# Quantification of Orbital Angular Momentum (OAM) beams states based of S system parameters.

Eldar Anufriyev eanufriyev@gmail.com

**Abstract-** In this paper, we propose a new quantitative method for determining the state number $\ell$ of the beam carry Orbital Angular Momentum (OAM), both individually and in a sum in a combination of $\ell$= -3, -2, -1, 0, 1, 2, 3 based on analysis of the S21 system parameters. This allows the receiving device to be tuned to receive beams with a specific OAM state. Radio waves carry OAM have a frequency of 80 GHz. The work is an experiment on a computer model. Simulation of beams with different states $\ell$, as well as determination of S system parameters, is carried out using the software product SCT Studio, and addition or superposition of S parameters of beams with different states $\ell$ was produced by means of the software product Matlab.

**Introduction.**

Recently, attention of scientific community was attracted by beams carry OAM, in connection with the prospect of their use for significant economy of the radio-frequency spectrum. OAM waves with different states don't interfere with one another [1], thus there can be, theoretically, unlimited number of overlaid beams on a single frequency for information transmission.

The need arises to determine beams with different OAM states $\ell$. Methods are given in the following papers that make it possible to distinguish beams with different $\ell$, both separately and in sum.

In [2] to discriminate between the two orthogonal OAM channels, transmitted on the same carrier frequency 2.4 GHz, they were frequency-modulated with constant level audio signals at different modulation frequencies (400 and 1000 Hz for $\ell$ =0 and $\ell$ =1, respectively). In a conventional single-antenna receiver setup, the two radio signals were heard simultaneously, and if two antennas are used and mechanically moved one relative to the other to select one of the orthogonal OAM beams, one signal is alternately suppressed with respect to the other due to the different spatial phase signature of two OAM states. This method has a number of disadvantages. Firstly, you need to mechanically move the antennas. Secondly, it is not possible to determine the sign of the orbital state by this method. Thirdly, it is not possible to determine a large number of beams with different OAM states.

The method based on graphical analysis of interferograms is used in [3] for the radio-frequency spectrum (28 GHz) and in [4], [5] for the optical spectrum. The state number $\ell$ of OAM beams can be deduced from the number of rotating arms in their interferograms, which are generated by combining the different OAM beams with a Gaussian beam ($\ell$=0) using a beam splitter. This method has the disadvantage that the only single beam with different $\ell$ can be determined, but it is impossible to determine sum of beams and the OAM state without addition with a Gaussian beam. In addition, this method lacks quantitative characteristics. Derivation on the number $\ell$ of the beam is made on the basis of a speculative conclusion about the number of arms and the direction of their twist.

In [6] in the optical range a pattern recognition algorithm (an artificial neural network) is used to identify the characteristic mode patterns displayed on a screen at the receiver. Authors of the article were able to distinguish between 16 different OAM mode superpositions with only a ~ 1,7% error rate and to use them to encode and transmit small grayscale images. To determine $\ell$, as in [3, 4, 5], analysis of superposition of fields is used. The disadvantage of this method is that it can not be fully applied to radio waves carry OAM, as well as applying of neural networks technology increases the processing time of signals.

In [7] for the radio-frequency spectrum, in [8] for the optical spectrum, a method for measuring the OAM spectrum of ultra-broadband radio and optical vortex pulses is proposed from fork like interferograms between vortex pulses and a reference plane wave pulse. It is based on spatial reconstruction, using the spatial Fourier transform, of the electric fields of the pulses to be measured from the frequency-resolved interference pattern. This method is demonstrated experimentally by obtaining the OAM spectra for different spectral components of the vortex pulses, which makes it possible to characterize the frequency dispersion of the topological charge of the OAM spectrum by a simple experimental setup. These articles also use the interferogram of one of the states $\ell$ with the Gaussian wave front, while it is impossible to determine several beams with different states $\ell$.

The aim of our article is to create univocal, accurate, reliable and simple method which allows determining OAM state $\ell$ of the beams using qualitative characteristics.

**Technique of the experiment.**

Our experiment is divided into two parts. In the first part, the system is simulated for the single state OAM $\ell$ using SCT Studio. The calculation of S21 parameters of receiving antennas is carried out. In the second part, using Matlab, the parameters S21 are interpolated, thus a pattern of the distribution of S21 parameters is obtained. Further, the sum of beams with different $\ell$ is simulated, and on the basis of an analysis of the distribution pattern of S21 parameters, it is concluded that there is a definite beam or a sum of them in various combinations.

**The first part of the experiment.**

To generate vortices with OAM, a pyramidal horn antenna and a spiral phase plate (SPP) are used, as in [9]. The pyramidal horn antenna, excited by a Gaussian pulse at a frequency fmin = 70 GHz fmax = 90 GHz, has the dimensions shown in Figure 1 and is made of pure copper. The SPP, which changes the Gaussian wave front by twisting it into a vortex, is made of Teflon and calculated by the method presented in [10] to generate vortices for six variants: $\ell$ = -3, -2, -1, +1, +2, + 3 for the frequency of 80 GHz. The step height of the spiral phase plate is 8.35 mm for $\ell$ = -1, +1, as shown in Figure 1, 16.7 mm for $\ell$ = -2, +2 -, 25.05 mm for $\ell$ = -3, +3. The SPP is not used for $\ell$=0. As receiving antennas symmetrical vibrators are used, the dimensions of which and relative position are shown in Figure 1 D. Receiving antennas are made of perfectly conductive material, presented in the CST Studio material table, to shorten the calculation time. The 13x13 matrix from the receiving antennas is located at a distance of 20 wavelengths, which is 75 mm, from the SPP when it is present or from the transmitting horn when $\ell$ = 0.

Seven experiments were performed with $\ell$ = -3, -2, -1, 0, +1, +2, +3, where S21 parameters at 80 GHz were calculated for each of the receiving antennas. Thus, seven 13x13 matrices were obtained, the cells of which contained S21 parameters of specific antennas.

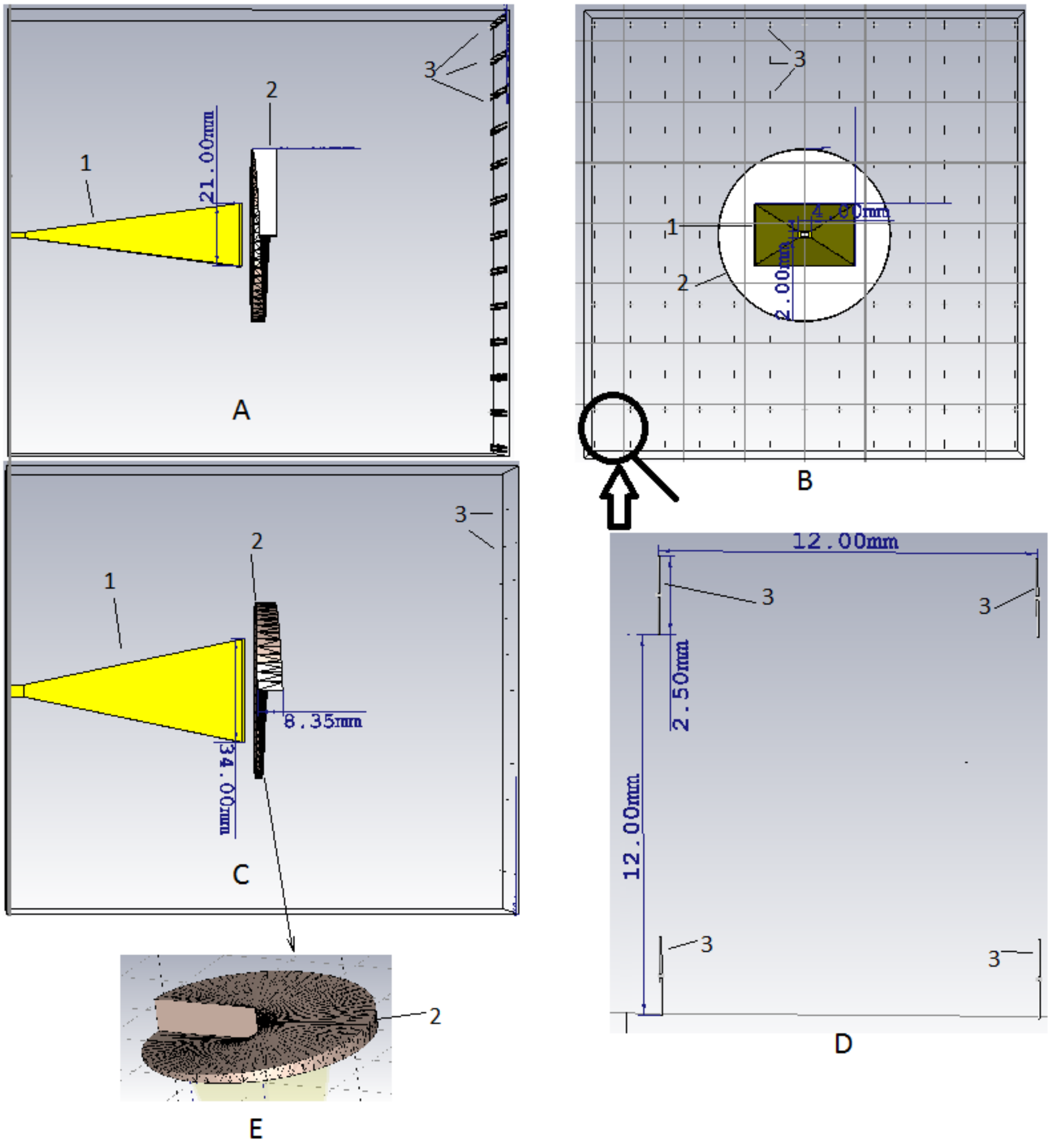

Figure 1. Simulation pattern of S21 parameters. A - main view, B - side view, C - top view, D - enlarged image of receiving antennas, E - spiral phase plate (SPP). Where 1 - horn transmission antenna, 2 - spiral phase plate, 3 - receiving antennas in the amount of 169.

Next, Matlab is used to analyze the table data.

**The second part of the experiment**.

To visualize the results, they were interpolated by cubic splines using Matlab, as shown in Figure 2.

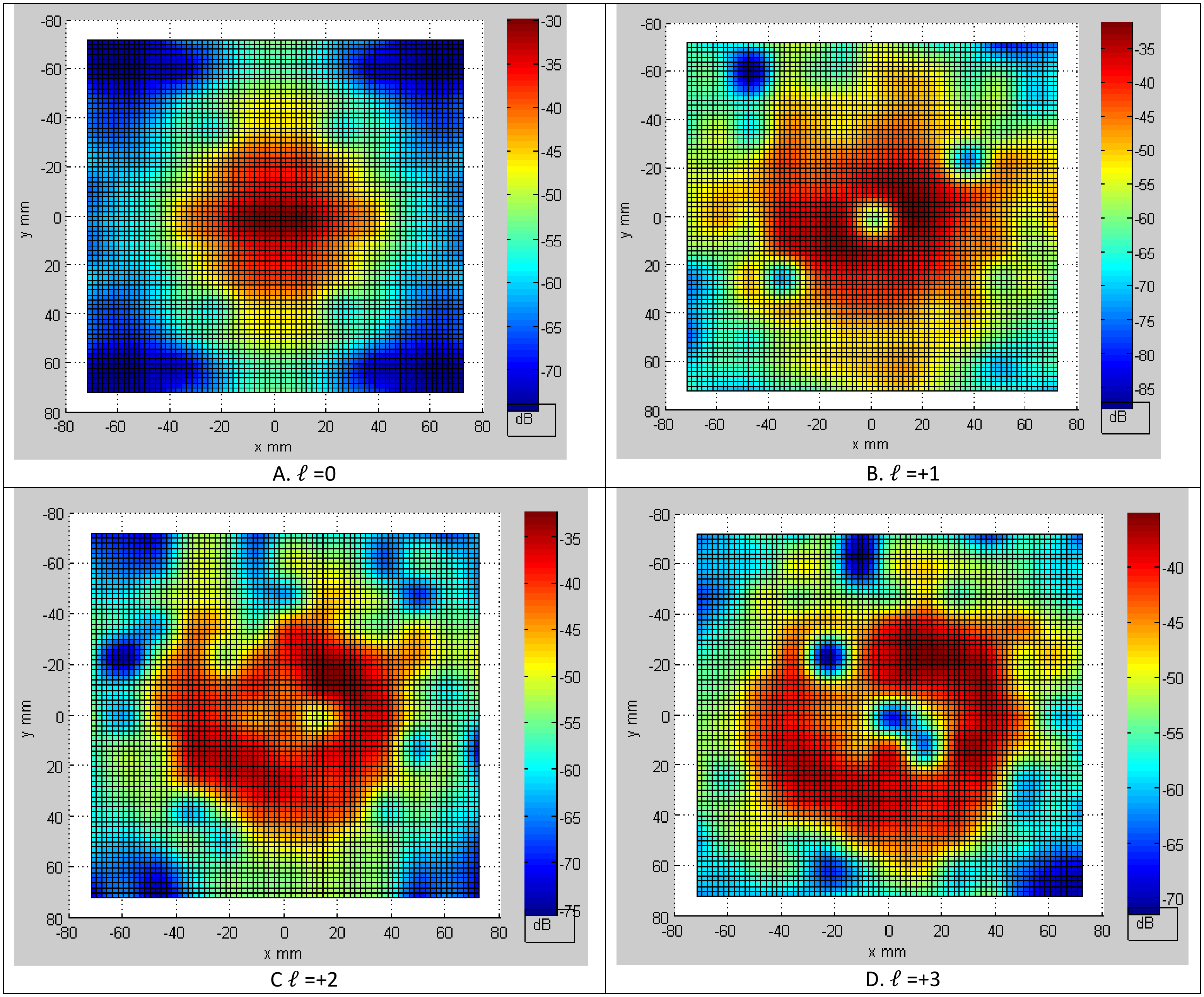

Figure 2. S21 parameters of receiving antennas interpolated by cubic splines.

The radius of a circle close to the peak, indicated by the dark red color, increases with increasing $\ell$. To assess whether this corresponds to real processes, we use the formula (10) from [11] for a beam with a single state $\ell$.

$$s(z)=w(z)\sqrt{\frac{N+1}{2}+|\ell_{min}|}\approx\frac{2z}{\kappa w_0}\sqrt{\frac{N+2|\ell_{min}|+1}{2}} \quad (10)$$

Where $\kappa$ is the wave number;

$w_0$ is beam waist radius;

$N = |\ell_{max}| - |\ell_{min}| + 1$ - amount of modes, from $\ell_{min}$ to $\ell_{max}$ all with the same sign;

Z is a distance from the SPP to the array of receiving antennas.

If there is one beam $\ell_{min}$ = $\ell_{max}$ =1, then the expression under the root in formula 10 is equal to 2, and if $\ell_{min}$ = $\ell_{max}$ =3, then the expression under the root in formula 10 will be 4. Thus, for the same $w_0$ with increasing $\ell$ the radius of a circle close to the peak grows, which is shown in Figure 2.

The idea of the proposed method is to quantify the beam with different OAM states by its characteristic maximum print. Under the characteristic maximum print is understood the position of maximus S21 parameters in the matrix of receiving antennas. It is necessary to select a level so that the beams with different $\ell$ have their own characteristic prints, and also sum of beams in different configurations would repeat the sum of their prints.

We choose a level in the gap between the maximum and the value corresponding to the difference between the maximum and 0.4 dB for positive beams. In the 13x13 matrix, all values equal to or greater than the maximum minus the selected level, are equal to one, the remaining cells are filled with zeros. The results of this experiment are summarized in Table 1 and presented in matrices 1-7.

The beams with $\ell$ = +1 и $\ell$ = +3 are summed up according to the formula (1) as if using beam splitters, as in [3]. The beam splitters are similar to those used in the optical mode. These beam splitters are realized by designing a matrix of reflective material on a dielectric substrate to obtain 50% transmission and 50% reflection at a frequency of 80 GHz. If more than two beams are added, then a cascade of these splitters is used.

$$S_{sum}21= 10\log(10^{0.1*S21(l=+1)} +10^{0.1S21(l=+3)}) \quad (1)$$

In Table 2, Green color indicates the axis of symmetry of the array of the receiving antennas. For clarity, the units are shown in purple.

The characteristic maximum print of beam with zero OAM contains the unit on the axes of symmetry, which is row 7 and line 7 (Matrix 1), when the characteristic maximum print of beam with $\ell$ =-1 has unit in row 8 and line 8 (Matrix 2). If there is interpolation of two beams with $\ell$ =+1, $\ell$ =+3, the characteristic maximum print of beams has 2 units – row 8 and line 6 and row 8 and line 5(Matrix 8).

It is possible to determine beams with positive OAM states, as well as their sum at the level 0.4 dB from the maximum. This level should be raised to 1.1 dB to indicate beams with negative and positive $\ell$, as well as their combinations.

Table 2. Characteristic maximum print of beams carry OAM.

Matrix 1. Global maximum at the level 0.4 dB. $\ell$ =0

| | | | | | | | | | | | | |
|---|---|---|---|---|---|---|---|---|---|---|---|---|
| 0 | 0 | 0 | 0 | 0 | 0 | 0 | 0 | 0 | 0 | 0 | 0 | 0 |
| 0 | 0 | 0 | 0 | 0 | 0 | 0 | 0 | 0 | 0 | 0 | 0 | 0 |
| 0 | 0 | 0 | 0 | 0 | 0 | 0 | 0 | 0 | 0 | 0 | 0 | 0 |
| 0 | 0 | 0 | 0 | 0 | 0 | 0 | 0 | 0 | 0 | 0 | 0 | 0 |
| 0 | 0 | 0 | 0 | 0 | 0 | 0 | 0 | 0 | 0 | 0 | 0 | 0 |
| 0 | 0 | 0 | 0 | 0 | 0 | 0 | 0 | 0 | 0 | 0 | 0 | 0 |
| 0 | 0 | 0 | 0 | 0 | 0 | 1 | 0 | 0 | 0 | 0 | 0 | 0 |
| 0 | 0 | 0 | 0 | 0 | 0 | 0 | 0 | 0 | 0 | 0 | 0 | 0 |
| 0 | 0 | 0 | 0 | 0 | 0 | 0 | 0 | 0 | 0 | 0 | 0 | 0 |
| 0 | 0 | 0 | 0 | 0 | 0 | 0 | 0 | 0 | 0 | 0 | 0 | 0 |
| 0 | 0 | 0 | 0 | 0 | 0 | 0 | 0 | 0 | 0 | 0 | 0 | 0 |
| 0 | 0 | 0 | 0 | 0 | 0 | 0 | 0 | 0 | 0 | 0 | 0 | 0 |
| 0 | 0 | 0 | 0 | 0 | 0 | 0 | 0 | 0 | 0 | 0 | 0 | 0 |

Matrix 2. Global maximum at the level 0.4 dB. $\ell$ =-1

| | | | | | | | | | | | | |
|---|---|---|---|---|---|---|---|---|---|---|---|---|
| 0 | 0 | 0 | 0 | 0 | 0 | 0 | 0 | 0 | 0 | 0 | 0 | 0 |
| 0 | 0 | 0 | 0 | 0 | 0 | 0 | 0 | 0 | 0 | 0 | 0 | 0 |
| 0 | 0 | 0 | 0 | 0 | 0 | 0 | 0 | 0 | 0 | 0 | 0 | 0 |
| 0 | 0 | 0 | 0 | 0 | 0 | 0 | 0 | 0 | 0 | 0 | 0 | 0 |
| 0 | 0 | 0 | 0 | 0 | 0 | 0 | 0 | 0 | 0 | 0 | 0 | 0 |
| 0 | 0 | 0 | 0 | 0 | 0 | 0 | 0 | 0 | 0 | 0 | 0 | 0 |
| 0 | 0 | 0 | 0 | 0 | 0 | 0 | 0 | 0 | 0 | 0 | 0 | 0 |
| 0 | 0 | 0 | 0 | 0 | 0 | 0 | 1 | 0 | 0 | 0 | 0 | 0 |
| 0 | 0 | 0 | 0 | 0 | 0 | 0 | 0 | 0 | 0 | 0 | 0 | 0 |
| 0 | 0 | 0 | 0 | 0 | 0 | 0 | 0 | 0 | 0 | 0 | 0 | 0 |
| 0 | 0 | 0 | 0 | 0 | 0 | 0 | 0 | 0 | 0 | 0 | 0 | 0 |
| 0 | 0 | 0 | 0 | 0 | 0 | 0 | 0 | 0 | 0 | 0 | 0 | 0 |
| 0 | 0 | 0 | 0 | 0 | 0 | 0 | 0 | 0 | 0 | 0 | 0 | 0 |

Matrix 3. Global maximum . $\ell$ =-2

| | | | | | | | | | | | | |
|---|---|---|---|---|---|---|---|---|---|---|---|---|
| 0 | 0 | 0 | 0 | 0 | 0 | 0 | 0 | 0 | 0 | 0 | 0 | 0 |
| 0 | 0 | 0 | 0 | 0 | 0 | 0 | 0 | 0 | 0 | 0 | 0 | 0 |
| 0 | 0 | 0 | 0 | 0 | 0 | 0 | 0 | 0 | 0 | 0 | 0 | 0 |
| 0 | 0 | 0 | 0 | 0 | 0 | 0 | 0 | 0 | 0 | 0 | 0 | 0 |
| 0 | 0 | 0 | 0 | 0 | 0 | 0 | 0 | 0 | 0 | 0 | 0 | 0 |
| 0 | 0 | 0 | 0 | 0 | 0 | 0 | 0 | 0 | 0 | 0 | 0 | 0 |
| 0 | 0 | 0 | 0 | 0 | 0 | 0 | 0 | 0 | 0 | 0 | 0 | 0 |
| 0 | 0 | 0 | 0 | 0 | 0 | 0 | 0 | 1 | 0 | 0 | 0 | 0 |
| 0 | 0 | 0 | 0 | 0 | 0 | 0 | 0 | 0 | 0 | 0 | 0 | 0 |
| 0 | 0 | 0 | 0 | 0 | 0 | 0 | 0 | 0 | 0 | 0 | 0 | 0 |
| 0 | 0 | 0 | 0 | 0 | 0 | 0 | 0 | 0 | 0 | 0 | 0 | 0 |
| 0 | 0 | 0 | 0 | 0 | 0 | 0 | 0 | 0 | 0 | 0 | 0 | 0 |
| 0 | 0 | 0 | 0 | 0 | 0 | 0 | 0 | 0 | 0 | 0 | 0 | 0 |

Matrix 4. Global maximum at the level 0.4 dB. $\ell$ =-3

| | | | | | | | | | | | | |
|---|---|---|---|---|---|---|---|---|---|---|---|---|
| 0 | 0 | 0 | 0 | 0 | 0 | 0 | 0 | 0 | 0 | 0 | 0 | 0 |
| 0 | 0 | 0 | 0 | 0 | 0 | 0 | 0 | 0 | 0 | 0 | 0 | 0 |
| 0 | 0 | 0 | 0 | 0 | 0 | 0 | 0 | 0 | 0 | 0 | 0 | 0 |
| 0 | 0 | 0 | 0 | 0 | 0 | 0 | 0 | 0 | 0 | 0 | 0 | 0 |
| 0 | 0 | 0 | 0 | 0 | 0 | 0 | 0 | 0 | 0 | 0 | 0 | 0 |
| 0 | 0 | 0 | 0 | 0 | 0 | 0 | 0 | 0 | 0 | 0 | 0 | 0 |
| 0 | 0 | 0 | 0 | 0 | 0 | 0 | 0 | 0 | 0 | 0 | 0 | 0 |
| 0 | 0 | 0 | 0 | 0 | 0 | 0 | 0 | 0 | 0 | 0 | 0 | 0 |
| 0 | 0 | 0 | 0 | 0 | 0 | 0 | 1 | 0 | 0 | 0 | 0 | 0 |
| 0 | 0 | 0 | 0 | 0 | 0 | 0 | 0 | 0 | 0 | 0 | 0 | 0 |
| 0 | 0 | 0 | 0 | 0 | 0 | 0 | 0 | 0 | 0 | 0 | 0 | 0 |
| 0 | 0 | 0 | 0 | 0 | 0 | 0 | 0 | 0 | 0 | 0 | 0 | 0 |
| 0 | 0 | 0 | 0 | 0 | 0 | 0 | 0 | 0 | 0 | 0 | 0 | 0 |

Matrix 5. Global maximum at the level 0.4 dB. $\ell$ =+1

| | | | | | | | | | | | | |
|---|---|---|---|---|---|---|---|---|---|---|---|---|
| 0 | 0 | 0 | 0 | 0 | 0 | 0 | 0 | 0 | 0 | 0 | 0 | 0 |
| 0 | 0 | 0 | 0 | 0 | 0 | 0 | 0 | 0 | 0 | 0 | 0 | 0 |
| 0 | 0 | 0 | 0 | 0 | 0 | 0 | 0 | 0 | 0 | 0 | 0 | 0 |
| 0 | 0 | 0 | 0 | 0 | 0 | 0 | 0 | 0 | 0 | 0 | 0 | 0 |
| 0 | 0 | 0 | 0 | 0 | 0 | 0 | 0 | 0 | 0 | 0 | 0 | 0 |
| 0 | 0 | 0 | 0 | 0 | 0 | 0 | 1 | 0 | 0 | 0 | 0 | 0 |
| 0 | 0 | 0 | 0 | 0 | 0 | 0 | 0 | 0 | 0 | 0 | 0 | 0 |
| 0 | 0 | 0 | 0 | 0 | 1 | 0 | 0 | 0 | 0 | 0 | 0 | 0 |
| 0 | 0 | 0 | 0 | 0 | 0 | 0 | 0 | 0 | 0 | 0 | 0 | 0 |
| 0 | 0 | 0 | 0 | 0 | 0 | 0 | 0 | 0 | 0 | 0 | 0 | 0 |
| 0 | 0 | 0 | 0 | 0 | 0 | 0 | 0 | 0 | 0 | 0 | 0 | 0 |
| 0 | 0 | 0 | 0 | 0 | 0 | 0 | 0 | 0 | 0 | 0 | 0 | 0 |
| 0 | 0 | 0 | 0 | 0 | 0 | 0 | 0 | 0 | 0 | 0 | 0 | 0 |

Matrix 6. Global maximum at the level 0.4 dB. $\ell$ =+2

| | | | | | | | | | | | | |
|---|---|---|---|---|---|---|---|---|---|---|---|---|
| 0 | 0 | 0 | 0 | 0 | 0 | 0 | 0 | 0 | 0 | 0 | 0 | 0 |
| 0 | 0 | 0 | 0 | 0 | 0 | 0 | 0 | 0 | 0 | 0 | 0 | 0 |
| 0 | 0 | 0 | 0 | 0 | 0 | 0 | 0 | 0 | 0 | 0 | 0 | 0 |
| 0 | 0 | 0 | 0 | 0 | 0 | 0 | 0 | 0 | 0 | 0 | 0 | 0 |
| 0 | 0 | 0 | 0 | 0 | 0 | 0 | 0 | 0 | 0 | 0 | 0 | 0 |
| 0 | 0 | 0 | 0 | 0 | 0 | 0 | 0 | 1 | 0 | 0 | 0 | 0 |
| 0 | 0 | 0 | 0 | 0 | 0 | 0 | 0 | 0 | 0 | 0 | 0 | 0 |
| 0 | 0 | 0 | 0 | 0 | 0 | 0 | 0 | 0 | 0 | 0 | 0 | 0 |
| 0 | 0 | 0 | 0 | 0 | 0 | 0 | 0 | 0 | 0 | 0 | 0 | 0 |
| 0 | 0 | 0 | 0 | 0 | 0 | 0 | 0 | 0 | 0 | 0 | 0 | 0 |
| 0 | 0 | 0 | 0 | 0 | 0 | 0 | 0 | 0 | 0 | 0 | 0 | 0 |
| 0 | 0 | 0 | 0 | 0 | 0 | 0 | 0 | 0 | 0 | 0 | 0 | 0 |
| 0 | 0 | 0 | 0 | 0 | 0 | 0 | 0 | 0 | 0 | 0 | 0 | 0 |

Matrix 7. Global maximum at the level 0.4 dB. $\ell$ =+3

| | | | | | | | | | | | | |
|---|---|---|---|---|---|---|---|---|---|---|---|---|
| 0 | 0 | 0 | 0 | 0 | 0 | 0 | 0 | 0 | 0 | 0 | 0 | 0 |
| 0 | 0 | 0 | 0 | 0 | 0 | 0 | 0 | 0 | 0 | 0 | 0 | 0 |
| 0 | 0 | 0 | 0 | 0 | 0 | 0 | 0 | 0 | 0 | 0 | 0 | 0 |
| 0 | 0 | 0 | 0 | 0 | 0 | 0 | 0 | 0 | 0 | 0 | 0 | 0 |
| 0 | 0 | 0 | 0 | 0 | 0 | 0 | 1 | 0 | 0 | 0 | 0 | 0 |
| 0 | 0 | 0 | 0 | 0 | 0 | 0 | 0 | 0 | 0 | 0 | 0 | 0 |
| 0 | 0 | 0 | 0 | 0 | 0 | 0 | 0 | 0 | 0 | 0 | 0 | 0 |
| 0 | 0 | 0 | 0 | 0 | 0 | 0 | 0 | 0 | 0 | 0 | 0 | 0 |
| 0 | 0 | 0 | 0 | 0 | 0 | 0 | 0 | 0 | 0 | 0 | 0 | 0 |
| 0 | 0 | 0 | 0 | 0 | 0 | 0 | 0 | 0 | 0 | 0 | 0 | 0 |
| 0 | 0 | 0 | 0 | 0 | 0 | 0 | 0 | 0 | 0 | 0 | 0 | 0 |
| 0 | 0 | 0 | 0 | 0 | 0 | 0 | 0 | 0 | 0 | 0 | 0 | 0 |
| 0 | 0 | 0 | 0 | 0 | 0 | 0 | 0 | 0 | 0 | 0 | 0 | 0 |

Matrix 8. Global maximum at the level 0.4 dB. Sum $\ell$ =+1, $\ell$ =+3

| | | | | | | | | | | | | |
|---|---|---|---|---|---|---|---|---|---|---|---|---|
| 0 | 0 | 0 | 0 | 0 | 0 | 0 | 0 | 0 | 0 | 0 | 0 | 0 |
| 0 | 0 | 0 | 0 | 0 | 0 | 0 | 0 | 0 | 0 | 0 | 0 | 0 |
| 0 | 0 | 0 | 0 | 0 | 0 | 0 | 0 | 0 | 0 | 0 | 0 | 0 |
| 0 | 0 | 0 | 0 | 0 | 0 | 0 | 0 | 0 | 0 | 0 | 0 | 0 |
| 0 | 0 | 0 | 0 | 0 | 0 | 0 | 1 | 0 | 0 | 0 | 0 | 0 |
| 0 | 0 | 0 | 0 | 0 | 0 | 0 | 1 | 0 | 0 | 0 | 0 | 0 |
| 0 | 0 | 0 | 0 | 0 | 0 | 0 | 0 | 0 | 0 | 0 | 0 | 0 |
| 0 | 0 | 0 | 0 | 0 | 1 | 0 | 0 | 0 | 0 | 0 | 0 | 0 |
| 0 | 0 | 0 | 0 | 0 | 0 | 0 | 0 | 0 | 0 | 0 | 0 | 0 |
| 0 | 0 | 0 | 0 | 0 | 0 | 0 | 0 | 0 | 0 | 0 | 0 | 0 |
| 0 | 0 | 0 | 0 | 0 | 0 | 0 | 0 | 0 | 0 | 0 | 0 | 0 |
| 0 | 0 | 0 | 0 | 0 | 0 | 0 | 0 | 0 | 0 | 0 | 0 | 0 |
| 0 | 0 | 0 | 0 | 0 | 0 | 0 | 0 | 0 | 0 | 0 | 0 | 0 |

Matrix 9. Global maximum at the level 0.4 dB. Sum ($\ell$ =+1, $\ell$ =+3) and $\ell$ =+2

| | | | | | | | | | | | | |
|---|---|---|---|---|---|---|---|---|---|---|---|---|
| 0 | 0 | 0 | 0 | 0 | 0 | 0 | 0 | 0 | 0 | 0 | 0 | 0 |
| 0 | 0 | 0 | 0 | 0 | 0 | 0 | 0 | 0 | 0 | 0 | 0 | 0 |
| 0 | 0 | 0 | 0 | 0 | 0 | 0 | 0 | 0 | 0 | 0 | 0 | 0 |
| 0 | 0 | 0 | 0 | 0 | 0 | 0 | 0 | 0 | 0 | 0 | 0 | 0 |
| 0 | 0 | 0 | 0 | 0 | 0 | 0 | 1 | 0 | 0 | 0 | 0 | 0 |
| 0 | 0 | 0 | 0 | 0 | 0 | 0 | 1 | 1 | 0 | 0 | 0 | 0 |
| 0 | 0 | 0 | 0 | 0 | 0 | 0 | 0 | 0 | 0 | 0 | 0 | 0 |
| 0 | 0 | 0 | 0 | 0 | 1 | 0 | 0 | 0 | 0 | 0 | 0 | 0 |
| 0 | 0 | 0 | 0 | 0 | 0 | 0 | 0 | 0 | 0 | 0 | 0 | 0 |
| 0 | 0 | 0 | 0 | 0 | 0 | 0 | 0 | 0 | 0 | 0 | 0 | 0 |
| 0 | 0 | 0 | 0 | 0 | 0 | 0 | 0 | 0 | 0 | 0 | 0 | 0 |
| 0 | 0 | 0 | 0 | 0 | 0 | 0 | 0 | 0 | 0 | 0 | 0 | 0 |
| 0 | 0 | 0 | 0 | 0 | 0 | 0 | 0 | 0 | 0 | 0 | 0 | 0 |

Matrix 10. Global maximum at the level 0.4 dB. Sum $\ell$ =-1, $\ell$ =-3

| | | | | | | | | | | | | |
|---|---|---|---|---|---|---|---|---|---|---|---|---|
| 0 | 0 | 0 | 0 | 0 | 0 | 0 | 0 | 0 | 0 | 0 | 0 | 0 |
| 0 | 0 | 0 | 0 | 0 | 0 | 0 | 0 | 0 | 0 | 0 | 0 | 0 |
| 0 | 0 | 0 | 0 | 0 | 0 | 0 | 0 | 0 | 0 | 0 | 0 | 0 |
| 0 | 0 | 0 | 0 | 0 | 0 | 0 | 0 | 0 | 0 | 0 | 0 | 0 |
| 0 | 0 | 0 | 0 | 0 | 0 | 0 | 0 | 0 | 0 | 0 | 0 | 0 |
| 0 | 0 | 0 | 0 | 0 | 0 | 0 | 0 | 0 | 0 | 0 | 0 | 0 |
| 0 | 0 | 0 | 0 | 0 | 0 | 0 | 0 | 0 | 0 | 0 | 0 | 0 |
| 0 | 0 | 0 | 0 | 0 | 0 | 0 | 1 | 0 | 0 | 0 | 0 | 0 |
| 0 | 0 | 0 | 0 | 0 | 0 | 0 | 1 | 0 | 0 | 0 | 0 | 0 |
| 0 | 0 | 0 | 0 | 0 | 0 | 0 | 0 | 0 | 0 | 0 | 0 | 0 |
| 0 | 0 | 0 | 0 | 0 | 0 | 0 | 0 | 0 | 0 | 0 | 0 | 0 |
| 0 | 0 | 0 | 0 | 0 | 0 | 0 | 0 | 0 | 0 | 0 | 0 | 0 |
| 0 | 0 | 0 | 0 | 0 | 0 | 0 | 0 | 0 | 0 | 0 | 0 | 0 |

Matrix 11. Global maximum at the level 0.4 dB. Sum ($\ell$ =-1, $\ell$ =-3) and $\ell$ =-2

| | | | | | | | | | | | | |
|---|---|---|---|---|---|---|---|---|---|---|---|---|
| 0 | 0 | 0 | 0 | 0 | 0 | 0 | 0 | 0 | 0 | 0 | 0 | 0 |
| 0 | 0 | 0 | 0 | 0 | 0 | 0 | 0 | 0 | 0 | 0 | 0 | 0 |
| 0 | 0 | 0 | 0 | 0 | 0 | 0 | 0 | 0 | 0 | 0 | 0 | 0 |
| 0 | 0 | 0 | 0 | 0 | 0 | 0 | 0 | 0 | 0 | 0 | 0 | 0 |
| 0 | 0 | 0 | 0 | 0 | 0 | 0 | 0 | 0 | 0 | 0 | 0 | 0 |
| 0 | 0 | 0 | 0 | 0 | 0 | 0 | 0 | 0 | 0 | 0 | 0 | 0 |
| 0 | 0 | 0 | 0 | 0 | 0 | 0 | 0 | 0 | 0 | 0 | 0 | 0 |
| 0 | 0 | 0 | 0 | 0 | 0 | 0 | 1 | 1 | 0 | 0 | 0 | 0 |
| 0 | 0 | 0 | 0 | 0 | 0 | 0 | 1 | 0 | 0 | 0 | 0 | 0 |
| 0 | 0 | 0 | 0 | 0 | 0 | 0 | 0 | 0 | 0 | 0 | 0 | 0 |
| 0 | 0 | 0 | 0 | 0 | 0 | 0 | 0 | 0 | 0 | 0 | 0 | 0 |
| 0 | 0 | 0 | 0 | 0 | 0 | 0 | 0 | 0 | 0 | 0 | 0 | 0 |
| 0 | 0 | 0 | 0 | 0 | 0 | 0 | 0 | 0 | 0 | 0 | 0 | 0 |

Matrix 12. Global maximum at the level 1.1 dB. Sum (($\ell$ =+1, $\ell$ =+3) and $\ell$ =+2) and (($\ell$ =-1, $\ell$ =-3) and $\ell$ =-2)

| | | | | | | | | | | | | |
|---|---|---|---|---|---|---|---|---|---|---|---|---|
| 0 | 0 | 0 | 0 | 0 | 0 | 0 | 0 | 0 | 0 | 0 | 0 | 0 |
| 0 | 0 | 0 | 0 | 0 | 0 | 0 | 0 | 0 | 0 | 0 | 0 | 0 |
| 0 | 0 | 0 | 0 | 0 | 0 | 0 | 0 | 0 | 0 | 0 | 0 | 0 |
| 0 | 0 | 0 | 0 | 0 | 0 | 0 | 0 | 0 | 0 | 0 | 0 | 0 |
| 0 | 0 | 0 | 0 | 0 | 0 | 0 | 1 | 0 | 0 | 0 | 0 | 0 |
| 0 | 0 | 0 | 0 | 0 | 0 | 0 | 1 | 1 | 0 | 0 | 0 | 0 |
| 0 | 0 | 0 | 0 | 0 | 0 | 0 | 0 | 0 | 0 | 0 | 0 | 0 |
| 0 | 0 | 0 | 0 | 0 | 0 | 0 | 1 | 1 | 0 | 0 | 0 | 0 |
| 0 | 0 | 0 | 0 | 0 | 0 | 0 | 1 | 0 | 0 | 0 | 0 | 0 |
| 0 | 0 | 0 | 0 | 0 | 0 | 0 | 0 | 0 | 0 | 0 | 0 | 0 |
| 0 | 0 | 0 | 0 | 0 | 0 | 0 | 0 | 0 | 0 | 0 | 0 | 0 |
| 0 | 0 | 0 | 0 | 0 | 0 | 0 | 0 | 0 | 0 | 0 | 0 | 0 |
| 0 | 0 | 0 | 0 | 0 | 0 | 0 | 0 | 0 | 0 | 0 | 0 | 0 |

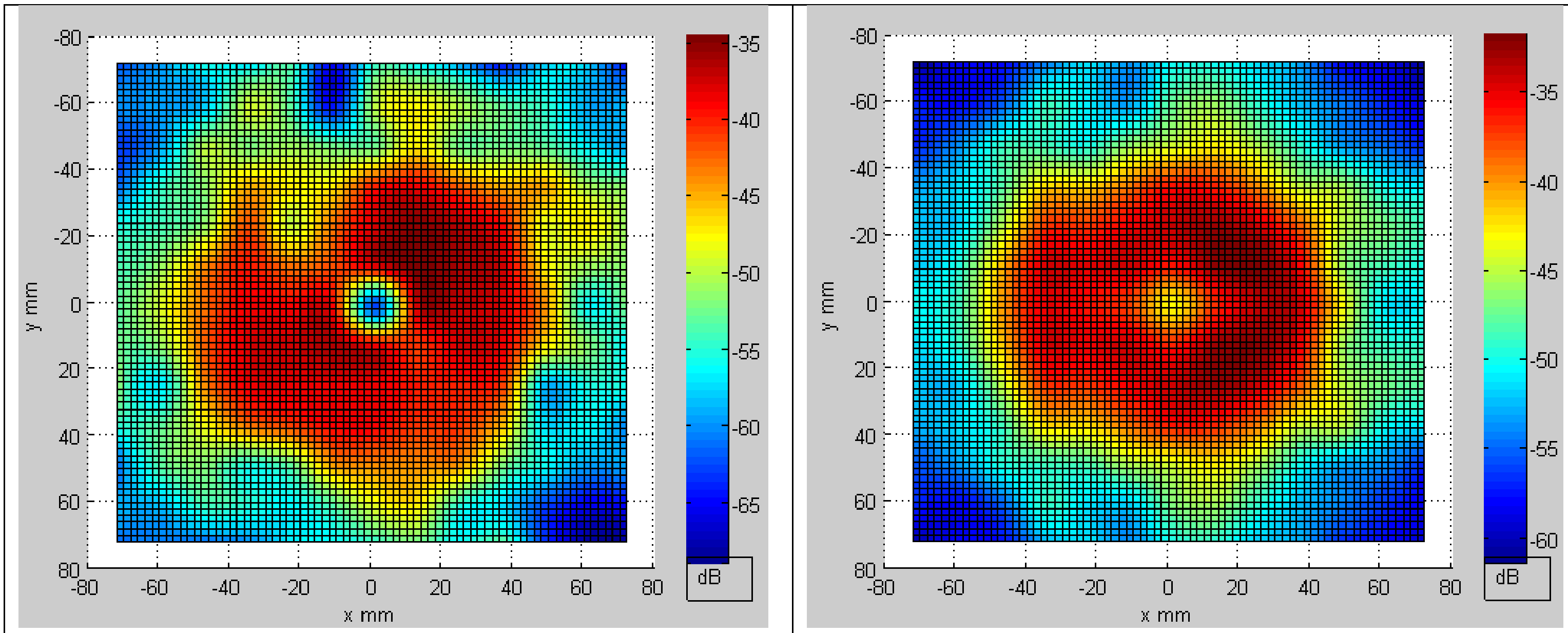

A ℓ =+1 and ℓ =+3

*B.* ((ℓ =+1, ℓ =+3) and ℓ =+2) and ((ℓ =-1, ℓ =-3) and ℓ =-2)

Figure 3. Interpolation of beams.

Figure 3A corresponds to Matrix 8. Dark red color indicates maximum of S21 parameters. It is important to decrease the maximum level to – 1.1 dB if we sum up beams with ((ℓ =+1, ℓ =+3) and ℓ =+2) and ((ℓ =-1, ℓ =-3) and ℓ =-2) because the maximum level 0.4 is not enough to indication. Figure 3B corresponds to Matrix 12.

**Conclusions. Discussion.**

We developed an accurate and simple method based on the qualitative characteristic maximum print for determination of states of Orbital Angular Momentum. With help of the matrix 13x13 of receiving antennas it is possible to predict six states: ℓ = -3, -2, -1, +1, +2, + 3 both singly and in sum. The repeatability of the experiment gives the same data, which indicates the reliability of our method.

**Acknowledgments**

I acknowledge Oleg Rutgaizer for interesting and fruitful discussion during this work. I thank Ruslan Anufriyev for computing power for calculations. I thank Timothy Victor for helpful comments.